# A combined experimental/numerical study on the scaling of impact strength and toughness in composite laminates for ballistic applications

Stefano Signetti[a,b], Federico Bosia[c], Seunghwa Ryu[b], Nicola M. Pugno[a,d,e*]

[a]*Laboratory of Bio-Inspired, Bionic, Nano, Meta Materials and Mechanics, Department of Civil, Environmental and Mechanical Engineering, University of Trento, via Masiano 77, I-38123 Trento, Italy*

[b]*Department of Mechanical Engineering, Korea Advanced Institute of Science and Technology, 291 Daehak-ro, Yuseong-gu, Daejeon 34141, Republic of Korea*

[c]*Department of Applied Science and Technology, Politecnico di Torino, corso Duca degli Abruzzi, 10129 Torino, Italy*

[d]*School of Engineering and Materials Science, Queen Mary University of London, Mile End Road E1 4NS, London, United Kingdom*

[e]*Ket-Lab, Edoardo Amaldi Foundation, via del Politecnico snc, I-0133, Roma, Italy*

*Corresponding author: nicola.pugno@unitn.it

**Abstract**

In this paper, the impact behaviour of composite laminates is investigated and their potential for ballistic protection assessed as a function of the reinforcing materials and structure for three representative fibre-reinforced epoxy systems involving carbon, glass, and para-aramid fibre reinforcements, respectively. A multiscale coupled experimental/numerical study on the composite material properties is performed, starting from single fibre, to fibre bundles (yarns), to single composite ply, and finally at laminate level. Uniaxial tensile tests on single fibres and fibre bundles are performed, and the results are used as input for non-linear finite element method (FEM) models for tensile and impact simulation on the composite laminates. Mechanical properties and energy dissipation of the single ply and multilayer laminates under quasi-static loading are preliminarily assessed starting from the mechanical properties of the constituents and subsequently verified using FEM simulations. FEM simulations of ballistic impact on multilayer armours are then performed, assessing the three different composites, showing good agreement with experimental tests in terms of impact energy absorption capabilities and deformation/failure behaviours. As result, a generalized multiscale version of the well-known Cuniff's criterion is provided as a scaling law, which allows to assess the ballistic performance of laminated composites, starting from the tensile mechanical properties of the fibres and fibre bundles and their volume fraction. The presented multiscale coupled experimental-numerical characterization confirms the reliability of the predictions for full scale laminate properties starting from the individual constituents at the single fibre scale.

## 1. Introduction

One of the main challenges in the development of protective armours against high-velocity impacts is to maximize the protection levels using lightweight materials and structures, since for many applications the use of large masses may be impractical or unsuitable, such as in aerospace applications. Conventional amours made with metal alloys or ceramic materials have been widely used in the past, with the latter guaranteeing comparable protection levels at almost a third of the weight of metals [1]. Amours made from these materials are isotropic, and their capability of stopping ballistic projectiles is proportional to the mass of the stopping materials, so that either the required minimum density or the thickness may become consistent for extreme protection levels. Therefore, these solutions are not applicable where low weight is fundamental to ensure unrestricted and efficient mobility, e.g. to terrestrial vehicles, aircraft, and spacecraft, or when material flexibility is desirable to guarantee ergonomics to body armour, such as for defense or sports applications [2].

In this regard, composite materials based on high performance fibre reinforcements exhibit high specific strength and stiffness [3, 4, 5], allowing the fabrication of relatively thin and flexible armours with good corrosion resistance [6]. Composites have good damage tolerance [6-7], superior fatigue properties [8], and excellent thermal and acoustic insulation [9]. They are also easy to fabricate, reducing costs and allowing flexibility in design [7, 9, 10], providing access to a combination of a wide range of materials that enable optimization for specific purposes. Another important characteristic is the limited degradation of properties after multiple impact events, i.e. the damage tolerance, which determines the long-term survivability of protective systems in harsh environments [6].

Armour protective capabilities are usually assessed in the terminal ballistic community on the basis of the so called $V_{50}$ parameter, i.e. the velocity corresponding to a 50% probability that the impacting mass is stopped by the target without perforation. According to the dimensional analysis carried out by Cuniff [11] for an elastic textile barrier constituted of fibres of density $\rho$, tensile strength $\sigma$, Young's modulus $E$, and failure strain $\varepsilon$, it was found that $V_{50} \sim U^{1/3}$, where $U = \frac{\sigma \varepsilon}{2\rho} \sqrt{\frac{E}{\rho}}$ is a parameter obtained as the product of the material-specific dissipated energy and the acoustic wave speed in the considered fibres. This acknowledged dimensional analysis allows to compare the actual protective performance of a wide range of fabrics. The advantage of employing composites over traditional metals and ceramics to increase the impact toughness clearly emerges due to their lower density, as well as higher strain to failure, specific strength and stiffness [3]. The good prediction capability of the above scaling criterion is an indication that fibre failure, both in tension and in shear due to shear plug [5], is one of the main damage mechanisms in multilayer composite armours, and it is then the primary source of energy absorption. Other principal damage mechanisms involve inter-layer delamination [12], matrix cracking and melting [13], fibre-matrix debonding, and fibre spallation [14]. However, the above mentioned criterion does not account for the real composites scenario where the volume fraction, different fibre orientation among different layers, and most of all size scale effects of material properties are not taken into account.

Reinforcing fibres employed in composites are usually assembled in unidirectional or bidirectional woven fabrics, in the form of dry preforms or pre-impregnated with resin (prepregs), in order to guarantee their uniform distribution and an even load transfer within the matrix [3, 15]. Short fibre reinforcements and random distributed long fibre mats are usually not suitable for ballistic applications due to their non-uniform microstructures. Woven fabrics are formed by interlacing two or more sets of bundle (yarns). Plain woven fabric is the simplest biaxial woven preform. More sets of yarns can also be used, and the resulting fabrics are called triaxial or multiaxial weaves, which progressively increase the grade of isotropy of mechanical properties of the fabric and of the resulting composite ply [16]. On the other hand, these architectures usually result in less compacted composite laminates with lower volume fractions with respect to bidirectional plain weaves, resulting in lower ballistic strength [17-18].

The complex mechanical behaviour emerging in laminate response, due to the presence of microscopic to macroscopic characteristic scales, requires a multiscale description from single fibre, to bundle, to ply and finally to laminate level for the selection of the optimal constituents and configurations for ballistic applications. Theoretical and computational methods can provide new insights in the comprehension of fracture mechanisms and of scaling of mechanical properties in heterogeneous/hierarchical/multiscale structures, beginning from microscale. One example for modelling fibrous materials is represented by so-called Hierarchical Fibre Bundle Models (HFBM) [19,20] where the mechanical properties of a fibre or thread at a given hierarchical level are statistically inferred from the average output deriving from repeated simulations at the lower level, down to the lowest hierarchical level, allowing the simulation of multiscale or hierarchical structures. Results show that specific hierarchical organizations can lead to increased damage resistance (e.g., self-similar fibre reinforced matrix materials) or that the interaction between hierarchy and material heterogeneity is critical, since homogeneous hierarchical bundles do not exhibit improved properties [21].

Moving up to the composite level, numerous theories have been proposed to date to describe the kinematics and stress states of composite laminates. Most of these laminate theories are extensions of the conventional, single-layer plate theories (e.g. Kirchoff-Love, Reissner-Mindlin [22, 23]) which are based on assumed variation of either stresses or displacements through the plate thickness. Equivalent Single Layer theories (ESL) [24-26] are simple extension of single layer theories accounting for variable sub-thickness and material properties in the solutions of partial differential equations of the single layer homogeneous plate. In carrying out the integration it is assumed that the layers are perfectly bonded. For many applications, the ESL theories provide a sufficiently accurate description of the global laminate response, e.g. tensile properties, transverse deflection, natural vibrations, critical buckling load. The main advantages of the ESL models are their inherent simplicity and low computational cost due to the relatively small number of variables. However, the ESL models are often inadequate for determining the three-dimensional stress field at the ply level, which may arise from severe bending or highly localized contact pressure. Moreover, the main shortcoming of the ESL models in modelling composite laminates is that the transverse strain components are continuous across interfaces between dissimilar (variable stiffness) materials. Unlike the ESL theories, layer-wise (or laminate shell) theories [27, 28] assume separate displacement field expansions within each material layer, thus providing a kinematically consistent representation of the strain field in discrete layer laminates, and allowing accurate determination of stresses within single plies. Such laminate theories are currently implemented in the most advanced element formulations in non-linear finite element method (FEM) codes [25].

Nowadays, these FEM approaches are capable of modelling the main mechanical phenomena which occur in high-velocity impact events such as contact, inter-layer delamination, material fracture and fragmentation, allowing the accurate replication of ballistic tests and their partial substitution in the design and optimization process [12, 29-31]. Such codes include sophisticated constitutive models, also accounting for strain-rate effects, and anisotropic failure criteria that allow the modelling of the most complex materials. In this regard, the accuracy and prediction capabilities in the design process of such models relies, at first, on the accurate characterization of material properties, which should be based on a multiscale approach. This would be fundamental, along the identification to key target parameters, for the application of machine learning techniques to optimize composites [32].

In this paper, we investigate the impact behaviour of three epoxy composite laminates reinforced with carbon, E-glass, and Twaron® (para-aramid, PA) fibres, respectively, and assess their potential for ballistic protection as a function of their structure and constitutive components. Up to the best of the authors' knowledge, the attempts so far, also very recent, have been limited to low velocity impacts and without a systematic investigation across all the dimensional scales involved [33, 34]. The aim is to create a simple multiscale characterization protocol that exploits the properties extracted from the single components at the microscale as input for reliable impact simulations at the macroscale. First, the tensile properties of single fibres and of the bundles (yarns) constituting the orthotropic woven textiles are characterized. Then, the obtained properties are used as input for FEM simulations to replicate tensile experiments on the laminates. Scaling of mechanical properties of interest with the characteristic sample size is also assessed. Finally, FEM impact simulations are performed to replicate experimental ballistic tests (initial projectile velocity $V_0 \cong 360$ m/s, impact kinetic energy $K_0 \cong 520$ J) on armours constituted by the previously characterized plies, computing their absorption capabilities and deformation/failure behaviour. The good agreement of impact simulation and ballistic experiments proves the validity of the proposed multiscale coupled experimental-simulation method. Finally, a multiscale generalization of the Cuniff's parameter is proposed to rationalize the results, providing a relatively simple scaling law that allows to assess and predict the ballistic performance of laminated composites, starting from the tensile mechanical properties of the fibres, their volume fraction and arrangement, which can provide preliminary design criteria with related time cost reductions in terms of prototyping and experimental tests.

## 2. Materials and methods

*2.1 Characterization of fibres, bundles and laminates*

We consider three of the most widely used fibre types in the manufacture of high-strength composites: carbon (T800), E-glass, and PA. The fibres were extracted from woven textile samples manufactured by G. Angeloni s.r.l, Italy, and commercially identified as fabrics *GG 301 T8*, *and VV-300 P, Style 286*, respectively.

First, single fibres were tested under uniaxial tension [35] using an Agilent T150 Nanotensile testing system (Figure 1.a), which allows sensitivity down to nN and nm on loads and displacements, respectively. 5 tests per fibre type were conducted. The samples, with a typical gauge length of 20 mm, were prepared in "C-shaped" paper frames and set-up in a clamped-clamped configuration in the sample holder (Figure 1.a). The paper frame is then cut and fibres loaded up to failure at a loading rate of 1 mm/min. The micro-fibres are analysed before and after testing using a Scanning Electron Microscope (SEM) to measure exact diameters. The fibre reinforcements for the considered laminated composites are in the form of fibre yarns in "plain weave" form, i.e. constituted by woven fibre bundles in mutually orthogonal directions ("weft" and "warp", see Figure S1 in the Supplementary Information). Given the non-uniformity in thickness and in density of the yarns and the uncertainty in the experimental determination of their actual thickness, the characterization is carried out on single fibre bundles with equivalent thickness properties. The experimental tests (5 tests per fibre type, due to high reproducibility of results) are performed after having measured the length $l$ (distance between clamps) and mass $m$ of the bundles and derived its cross-section area as $A = m/(l\rho_i)$, where $\rho_i$ is the volumetric density of the corresponding material. The determined bundle cross-section areas are consistent with the values derived from the ratio between the linear density provided by the producers' specifications (in dtex, $= 10^{-7}$ kg/m) and the known volumetric density of the materials. Force-displacement $(F - \delta)$ curves are measured using a MTS uniaxial testing system (with a 1 kN load cell, Figure 1.b), and converted to stress ($\sigma = F/A$) strain ($\varepsilon = \delta/l$). From these overall properties are derived, such as Young's modulus $E = \sigma/\varepsilon$, fracture strength ($\sigma_F = \max\{\sigma\}$), and ultimate strain ($\varepsilon_F = \max\{\varepsilon\}$). The load application velocity is 1 mm/min.

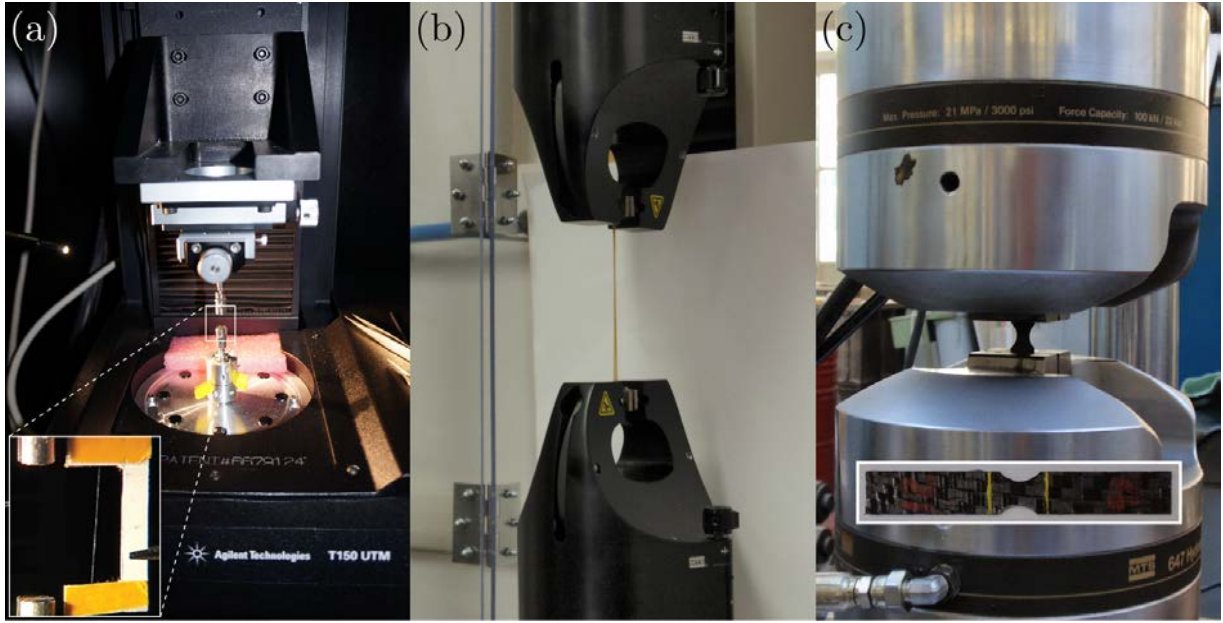

**Figure 1.** Materials characterization from micro- to the laminate scale. (a) Microtensile characterization of single fibres (the inset illustrates the "C-shaped" frame for placement of the fibre in the loading cell); (b) mesoscale characterization of fibre bundles extracted from the textiles; (c) macrotensile characterization of laminates (a typical carbon sample is shown in the inset).

Mechanical tests are also performed on laminated composite specimens [36] fabricated by Vemar s.r.l, Italy with the above textiles. The used resin is a thermoset *Bakelite® EPR L 1000 – set* by Bakelite AG. Single ply, 5-ply and 10-ply specimens are considered, with 0° and 45° orientation of the textile warp with respect to the loading direction. The different thicknesses and fibre orientations are considered to provide data for general conclusions, independent of the specific considered geometry. Specimen dimensions have a length of 10 cm and width of 15 mm. Small circular cuts (9 mm radius) are performed in the central part of the samples to prepare dog-bone specimens (Figure 1.c and Supplementary Figure S2). Four specimen for each material/thickness/orientation subgroups are performed. Average thickness and volume fractions of the single and multilayer plies and are reported in Table 1 (see Table S1 in the Supplementary information for more extended dimensional characteristics). The tests are performed using another MTS uniaxial testing system, with a 10 kN load cell (Figure 1.c) and a loading rate of 1mm/min.

**Table 1.** Average ply thickness *t* and fibre volume fraction *f* (and related standard deviation) of the carbon, E-glass, and PA fibre-based composites with 1, 5, 10 layers at 0° and 45° orientation with respect to the direction of load application. The volume fraction *f* is determined assuming average textile thickness of 0.12 mm, 0.12 mm, and 0.10 mm for carbon, E-glass, and PA woven textiles, respectively, as specified from the producers. Data for all tested samples for each category are reported in full in Table S1 in the Supplementary Information.

|  | 0° | | | | | | 45° | | | | | |
|---|---|---|---|---|---|---|---|---|---|---|---|---|
|  | 1 layer | | 5 layers | | 10 layers | | 1 layers | | 5 layer | | 10 layers | |
|  | *t* [mm] | *f* | *t* [mm] | *f* | *t* [mm] | *f* | *t* [mm] | *f* | *t* [mm] | *f* | *t* [mm] | *f* |
| **Carbon** | 0.278 ± 0.0083 | 0.432 ± 0.0128 | 0.254 ± 0.218 | 0.473 ± 0.0381 | 0.270 ± 0.0187 | 0.444 ± 0.0326 | 0.298 ± 0.0083 | 0.403 ± 0.0111 | 0.257 ± 0.1139 | 0.468 ± 0.0392 | 0.235 ± 0.0500 | 0.511 ± 0.0109 |
| **Glass** | 0.275 ± 0.0433 | 0.406 ± 0.0866 | 0.185 ± 0.0087 | 0.649 ± 0.0289 | 0.178 ± 0.0043 | 0.676 ± 0.0170 | 0.325 ± 0.0433 | 0.369 ± 0.0433 | 0.240 ± 0.2121 | 0.500 ± 0.0874 | 0.181 ± 0.0249 | 0.664 ± 0.0092 |
| **PA** | 0.225 ± 0.0433 | 0.533 ± 0.0722 | 0.185 ± 0.0087 | 0.649 ± 0.0241 | 0.263 ± 0.0083 | 0.457 ± 0.0122 | 0.300 ± 0.0707 | 0.400 ± 0.0908 | 0.175 ± 0.0433 | 0.686 ± 0.0301 | 0.243 ± 0.1090 | 0.495 ± 0.0181 |

*2.2 FEM tensile simulations*

Dog-bone shaped laminate samples (geometrical characteristics in Figure S2 in the Supplementary Information) have been reproduced by FEM model to evaluate the capability of numerically capturing the elastic and fracture behaviour of laminates and comparing them with the results of experimental measurements with approximate predictions by a rule of mixtures. The LS-DYNA® v971 R10.1 solver by Livermore Software Technology Corporation (LSTC) [37] was used in this study. 8 node solid-shell (also "thick-shell") elements based on the Reissner-Mindlin kinematic assumption [22, 23] and developed by Liu et al. [38] were employed for the simulations. This element formulation (TSHELL ELFORM=1 [37]) allows the implementation of the laminate shell theories for an accurate computation of transversal stresses within the ply. A single point reduced in-plane integration rule was adopted. Although higher order in-plane integration schemes, e.g. 2x2 Gauss quadrature, could be chosen, we opted for this formulation since low-order integration schemes are the most robust when element become largely distorted, as may happen in high-velocity impact simulations. Thus, we opted to use the same formulation in tensile simulation tests as that used in the more critical impact simulations presented later. Since single point quadrature is related to a reduction of the stiffness matrix, spurious zero-energy modes of deformation (also known as *hourglass* modes) may arise, as usually occurs under concentrated pressures. A viscous form hourglass control [37], i.e. introducing a fictitious viscosity, was used in the simulations (LS-DYNA hourglass type 3 [35]). We checked the fictitious energy introduced to mitigate hourglassing to be below 5% of the deformation energy at each simulation time for the whole model and for each of its deformable subparts (single plies). The ply thickness and volume fraction associated with each of the simulated cases were determined according to the measurements on experimental laminates (see Table 1). One single element through the thickness was used to model the single plies. Given the variable thickness of the plies of the various tested specimens (Table 1), the aspect ratios for the elements in the notched part of the specimens vary in ranges from ~1:1:0.68 ($x$, $y$, $z$) to ~1:1:1.25, with an in plane characteristic size of about 0.26 mm (see Figure S2 in the Supplementary Information), as results from the performed convergence study (see Section S2.1 in the Supplementary Information). The thick shell element was sampled with 14 integration points (IPs) through the thickness, of which the 6 innermost were associated to the core of woven textile, while the outermost (4+4) were attributed to the epoxy matrix. The resulting integration scheme for all 18 simulated laminates is summarized in Table S3 in the Supplementary Information. MAT 58 (LAMINATED_COMPOSITE_FABRIC [37]) was used to simulate the fabric materials. This is a continuum damage model based on the Matzenmiller-Lubliner-Taylor theory [40] intended to describe the failure of woven fabrics and composite laminates, also accounting for post-critical behaviour. More details about the model are reported in the Supplementary Information, Section 2.2 and the input parameters for carbon, glass, and PA fibres, as extracted from our experiments, and for the epoxy resin (as specified by the producer) are reported in Supplementary Tables S5-S8. Average values of thickness and volume fraction reported in Table 1 were used. Thus, in total 18 simulation were performed corresponding to single cases determined by material, number of layers, and orientation of the textile with respect to the application of the load.

*2.3 FEM impact simulations*

Four armours, based on the characterized materials and corresponding to the conducted experimental test, were simulated: a 17-layer carbon-based armour with overall thickness of 4 mm ($f = 0.510$), a 16-layer glass-based armour with a thickness of 3 mm ($f = 0.640$), and two 30-layer PA armours with thicknesses of 5 and 7 mm ($f = 0.599$ and $f = 0.429$, respectively). The integration scheme, element formulation, and material model follow the same setup adopted for the tensile testing simulations. The integration scheme for the four tested armours is reported in detail in Table S4 in the Supplementary Information. The simulated target is comprised of a circular plate (only one quarter is simulated due to the symmetry of the system) subjected to the impact of a lead/copper projectile simulating a FMJ Remington 9 mm Parabellum (diameter $\phi$ = 9.02 mm and mass $m_P$ = 8.04 g) traveling at 360 m/s (Figure 2), i.e. resulting in an impact energy of about 520 J. The plate radius is $R$ = 40 mm, which is about 9 times larger than the radius of the projectile, so that edge effects can be neglected, and the plate is fully clamped at the external edge. The woven orientation from each layer to the next progressively increases by an angle of 45° (i.e. stacking sequence: $k$ [0°, 45°, 90°, -45°]). One single element through the thickness was used to model the single plies. Given the variable thickness of the plies of the various targets, the aspect ratios for the elements in the region under impact (< 3$R$) are in the range ~1:1:0.42 ($x$, $y$, $z$) to ~1:1:0.59, with an in plane characteristic size of about 0.40 mm (see details of the mesh in Figure S4 in the Supplementary Information) as results from the performed convergence study (see Section S2.1 in the Supplementary Information).

An eroding type segment-to-segment contact is implemented between the plate layers (static and dynamic coefficient of friction equal to $\mu_S$ = 0.20, $\mu_D$ = 0.15, respectively [41]). A stress-based segment-to-segment tiebreak type contact (LS-DYNA - Option 6) was implemented to model inter-layer adhesion and delamination with normal and shear limit stresses equal to NFLS = 0.35 GPa and SFLS = 0.10 GPa, respectively. Finally, a segment-to-segment contact is implemented between the projectile and the target layers ($\mu_S$ = 0.40, $\mu_D$ = 0.30 [41]): in this case the `SOFT=2` option was activated to avoid interpenetration, given the high mismatch between the projectile and the composite contact stiffnesses. No scaling of the contact stiffness of the slave/master surfaces was operated in the contact. More details of the contact implementation can be found in the Supplementary Information, Section 2.4 where the script lines regarding contact implementation are also reported (Tables S9-S11). Failure within the armour is implemented by means of element erosion, which is based on the failure criterion of the specific material model (MAT_58 [40]): when failure is reached at all the integration points the element is deleted from the simulation, properly accounting for its energy in the overall balance. Again, hourglass energy was verified to be less than 5% of the deformation energy at each simulation timestep for the whole model and for each of its deformable parts separately. The total simulation time for all simulations was of 0.07 ms, which ensured complete stop or penetration of the target with stabilization of the projectile residual velocity ($V_{res}$).

In this work, strain-rate effects on material properties, although generally important for this problems, are not considered. First, the size-scaling of material properties, rather then they absolute magnitude, would be in very limited part affected. Secondly, the dimensional analysis of interest of this work for the impact behavior through the Cuniff's criterion ($V_{50}/U^{1/3}$) conceptually eliminates the strain-rate dependency of results.

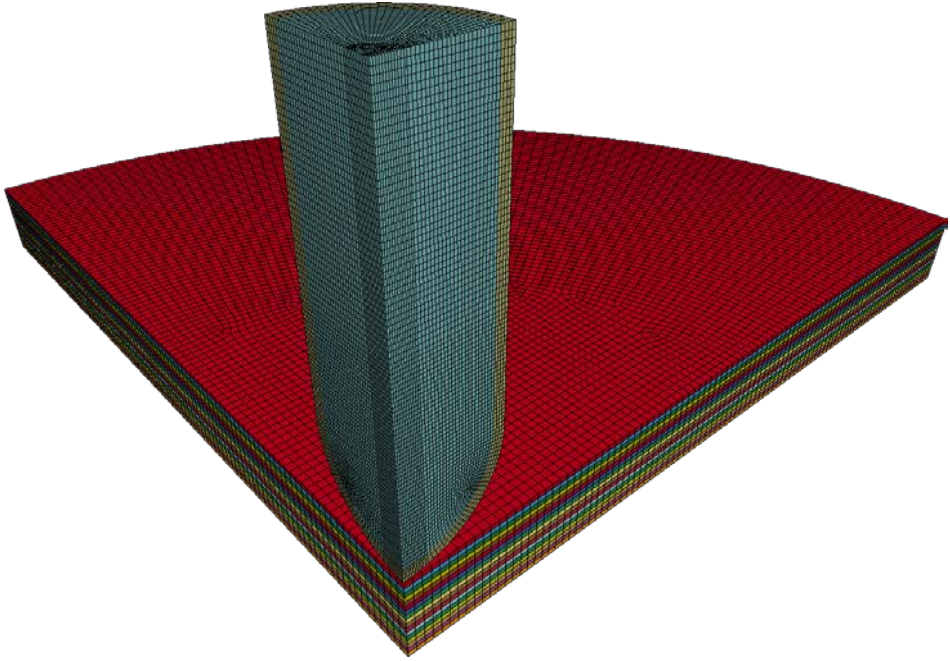

**Figure 2.** Finite element model for impact simulation (Carbon T800 17-layer laminate).

## 3. Results and discussion

*3.1 Characterization of single fibres and fibre bundles*

Typical results for mechanical microtensile tests and fibre volumetric characterization are summarized in Table 2. The fibres display approximately a linear stress-strain behaviour up to failure, which occurs between 1% and 3.1% strain, and between 1.24 and 4.17 GPa stress, with glass fibres displaying a considerably smaller strength, carbon displaying the maximum strength and PA the maximum toughness (integral of the force-displacement curve divided by the fibre mass). The results fall within the reported range in existing literature [3, 33, 42]. PA also displays the largest Cuniff's parameter, and is thus expected to be the most suitable material for energy dissipation by material failure.

Typical stress-strain results for various fibre bundle samples are shown in Figure 3, and the extracted mechanical parameters reported in Table 3. In general, tests on fibre bundles yield smaller strength values compared to single fibres (Figure 4). This can be attributed to the statistical distributions in the strength and in the ultimate strain of the single fibres, leading to a non-simultaneous breaking of the fibres (Figure 3), as predicted by HFBM [19, 20]. This is demonstrated by the various peaks in the stress-strain curves, and a maximum stress reached for a given percentage of surviving fibres (Figure 3). This type of mechanical test provides a more reliable estimation of the properties of the fibre yarns in the composites, and thus we used these values in the numerical simulations. Using classical Weibull's statistic [43] to study the distribution of the fracture strength of bundles under uniaxial uniform stress, we have:

$$F(\sigma_i) = 1 - \exp\left[-\frac{A_i}{A_0}\left(\frac{\sigma_i}{\sigma_{A,0}}\right)^{1/m}\right] \qquad (1)$$

where $\sigma_{A,0}$ and $m$ are the Weibull's shape and scale parameters, respectively, for a specific set of samples (material) and $F(\sigma_i) = \left(i - \frac{1}{2}\right)/N$ is the probability of failure of the $N$ samples sorted in order of increasing strength [44] (data in Table S12 in the Supplementary Information). $A_0$ and $\sigma_{A,0}$ are in our case the average values of the cross-section area and of the tensile strength, respectively, of the single fibre of the considered material (determined from diameter and strength values, respectively, reported in Table 2). For the studied materials we determine $m$ to be 9.4, 29.9, 26.8 for carbon, E-glass, and PA fibres, respectively. The quasi-linear behaviour up to fracture in PA bundle stress-strain curves implies that there is small dispersion on the strength values of the single fibres, contrary to the carbon and E-glass cases (Figure 3), as also quantified by the Weibull analysis. Carbon fibres display high strength values but fragile fracture and dispersion in strength values, which may lead to low fracture toughness, and therefore limited impact strength. PA fibres, on the other hand, exhibit good strength characteristics with greater toughness values. Finally, E-Glass yarns have smaller strength values as compared to carbon and PA. Our Weibull's analysis is qualitatively and quantitatively in agreement with other results of systematic studies recently published [42], showing the consistent measurement of properties at the fibre and bundle scale.

**Table 2.** Average tensile mechanical and volumetric properties of the single fibres.

|  | **Carbon** | **E-Glass** | **PA** |
|---|---|---|---|
| **Young's modulus [GPa]** | 232.77±19.6 | 55.11±20.2 | 95.27±9.7 |
| **Strength [GPa]** | 4.12±0.7 | 1.24±0.4 | 2.82±0.4 |
| **Ultimate strain** | 0.018±0.004 | 0.023±0.007 | 0.030±0.001 |
| **Toughness (av.) [J/m$^3$]** | 0.0365 | 0.0140 | 0.0417 |
| **Diameter [µm]** | 6 | 20 | 12 |
| **Density [kg/m$^3$]** | 1810 | 2540 | 1445 |
| **U$^{1/3}$ (av.) [m/s]** | 611 | 300 | 617 |

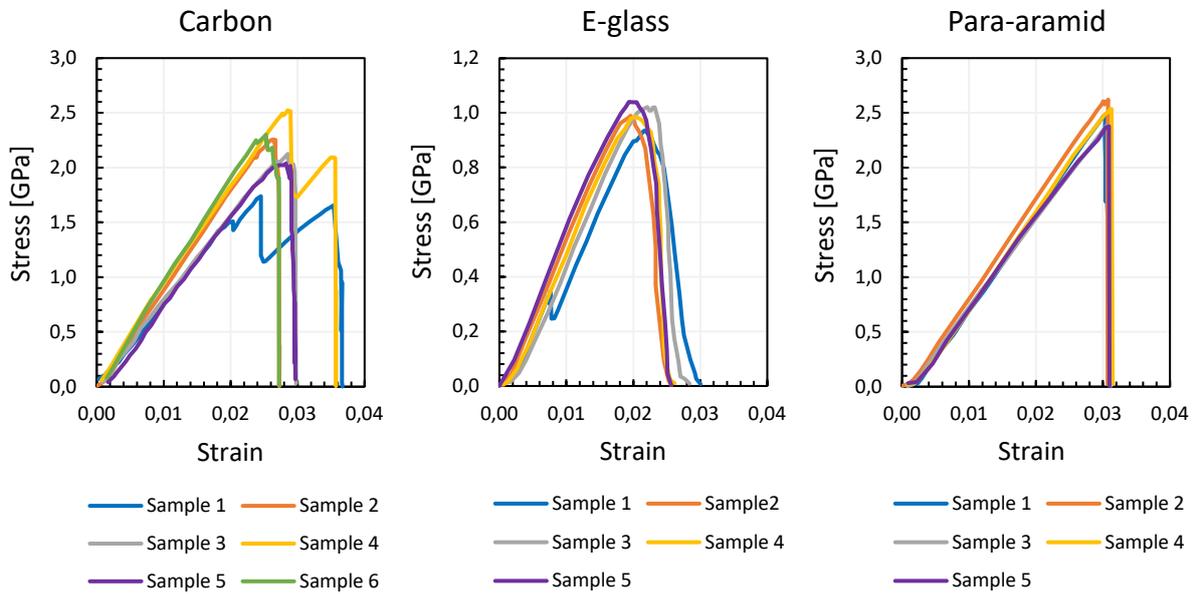

**Figure 3.** Experimental tensile stress-strain curves for various samples of carbon, E-glass and PA fibre bundles tested up to failure.

**Table 3.** Average tensile mechanical and volumetric properties of the fibre bundles. Values for all tested samples for Weibull analysis are reported in Table S12 in the Supplementary Information.

|  | **Carbon** | **E-Glass** | **PA** |
|---|---|---|---|
| **Young's modulus [GPa]** | 85.7±10.13 | 48.32±8.68 | 72.94±4.14 |
| **Strength [GPa]** | 2.17±0.27 | 0.995±0.04 | 2.52±0.09 |
| **Strain at peak** | 0.026±0.003 | 0.021±0.001 | 0.031±0.001 |
| **Ultimate strain** | 0.031±0.004 | 0.027±0.002 | 0.031±0.001 |
| **Area [mm$^2$]** | 0.255 (0.246*) | 0.113 (0.118*) | 0.092 (0.084*) |
| **Weibull parameter *m*** | 9.4 | 29.9 | 26.8 |

*Value obtained as ratio of the linear density of the textile in [dtex], as declared by the producers, and the volumetric bulk density of the material in [kg/m$^3$], is included for validation of performed measure.

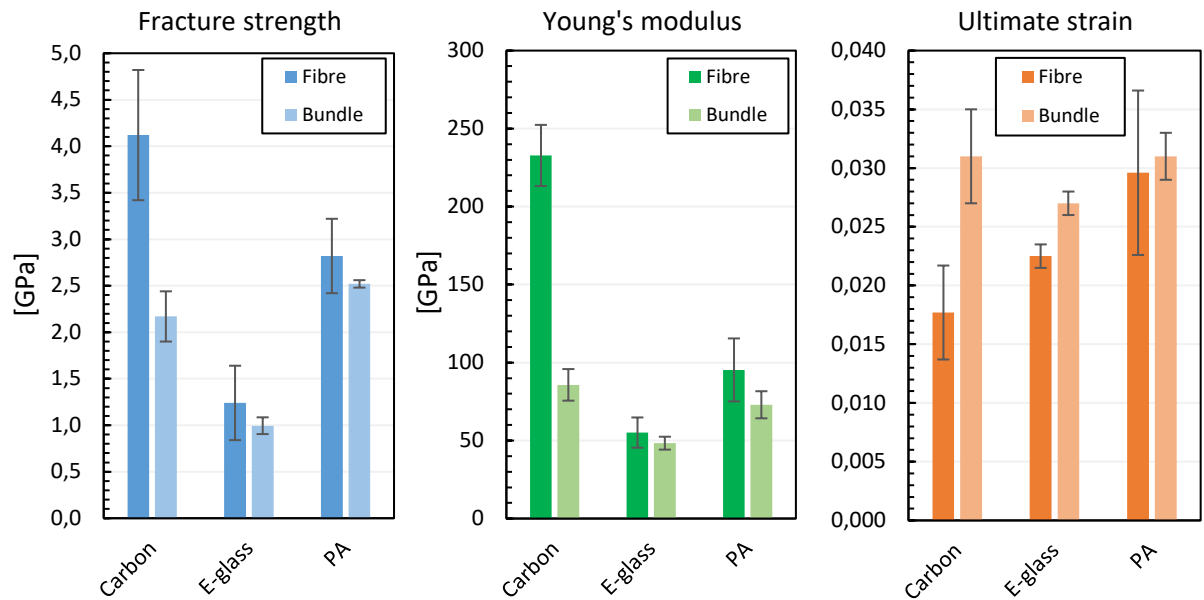

**Figure 4.** Comparison of average strength, Young's modulus, and ultimate strain (and related standard deviations) for Carbon, E-glass and PA fibres and corresponding bundles.

*3.2 Scaling of laminate properties*

Results for uniaxial tests on dog-bone specimens for 1-ply, 5-ply and 10-ply laminates are summarized in Figure 5. These results are compatible with those commonly found in literature for the considered materials [45, 46]. As an example case, the resulting experimental stress-strain curves for 1-ply PA laminates (0° and 45° woven direction), together with the results of the numerical simulations, are reported in Figure 6 (for the experimental and simulation-derived stress-strain curves of all other materials and laminates with different number of plies see Figures S5-S12 in the Supplementary Information).

In Figure 5 we can observe from simulation values that, generally, the strength diminishes with the increase of the number of layers according to well-known size effects on fracture properties. This trend is occasionally inverted due to the fact that the tested experimental samples, and thus the simulated counterparts, are not compared over an equal volume fraction basis, originated by the production process. Indeed, a clear dependence of strength on the volume fraction is observed (especially in PA and E-glass laminates in Figure 5). Simulation and experiment are generally in good agreement, although a significant variance in the experimental results is observed, especially in the 1-ply Carbon based samples, probably due to residual defects from manufacturing. This underlines the importance of the production process in providing final composite with actual predicted mechanical properties from its constituents and a sufficient and reliable level of performance, as well as the employment of reliable simulation models when few characterization tests are available. The in-plane fracture strength of the laminates ($\sigma_c = \max\{\sigma\}$), with different orientation $\theta$ of the woven with respect to the direction of application of the load, can be derived from the fracture strength of the fibres ($\sigma_f$) and matrix ($\sigma_m$) by application of the following rule of mixture which takes in to account the orthotropic nature of the woven textile:

$$\begin{Bmatrix}\sigma_{c,x}\\ \sigma_{c,y}\end{Bmatrix} = f \begin{Bmatrix}\sigma_{f,1}\\ \sigma_{f,2}\end{Bmatrix} \begin{bmatrix} cos^4\theta & sin^4\theta \\ \sin^4\left(\theta + \frac{\pi}{2}\right) & \cos^4\left(\theta + \frac{\pi}{2}\right)\end{bmatrix} + (1-f)\begin{Bmatrix}\sigma_{m,1}\\ \sigma_{m,2}\end{Bmatrix} \qquad (2)$$

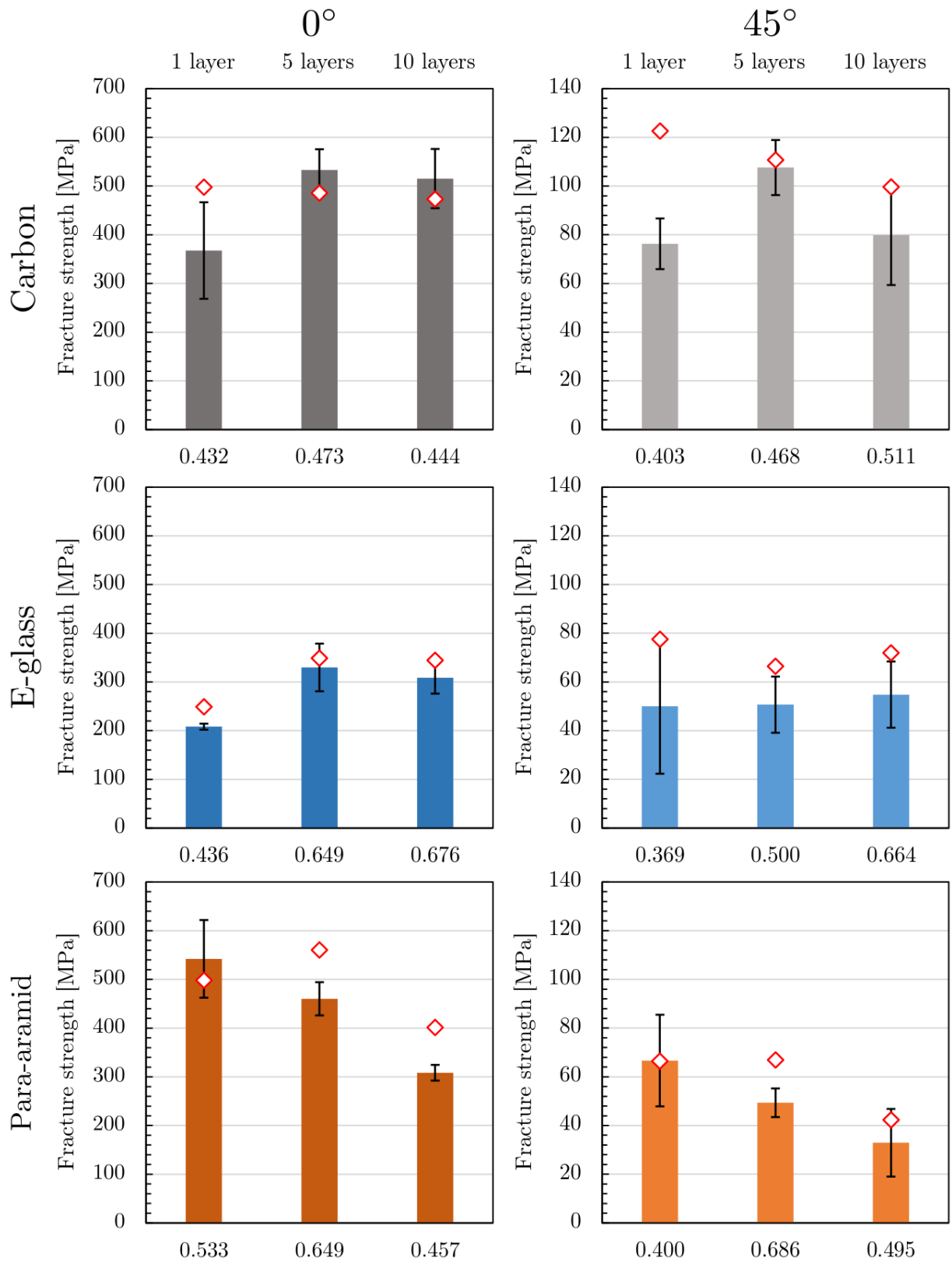

**Figure 5.** Laminate tensile strength from experimental data (columns representing mean, standard deviation is reproduced by bars) and comparison with FEM simulation results (red dots). See Table 4 for the corresponding numerical values.

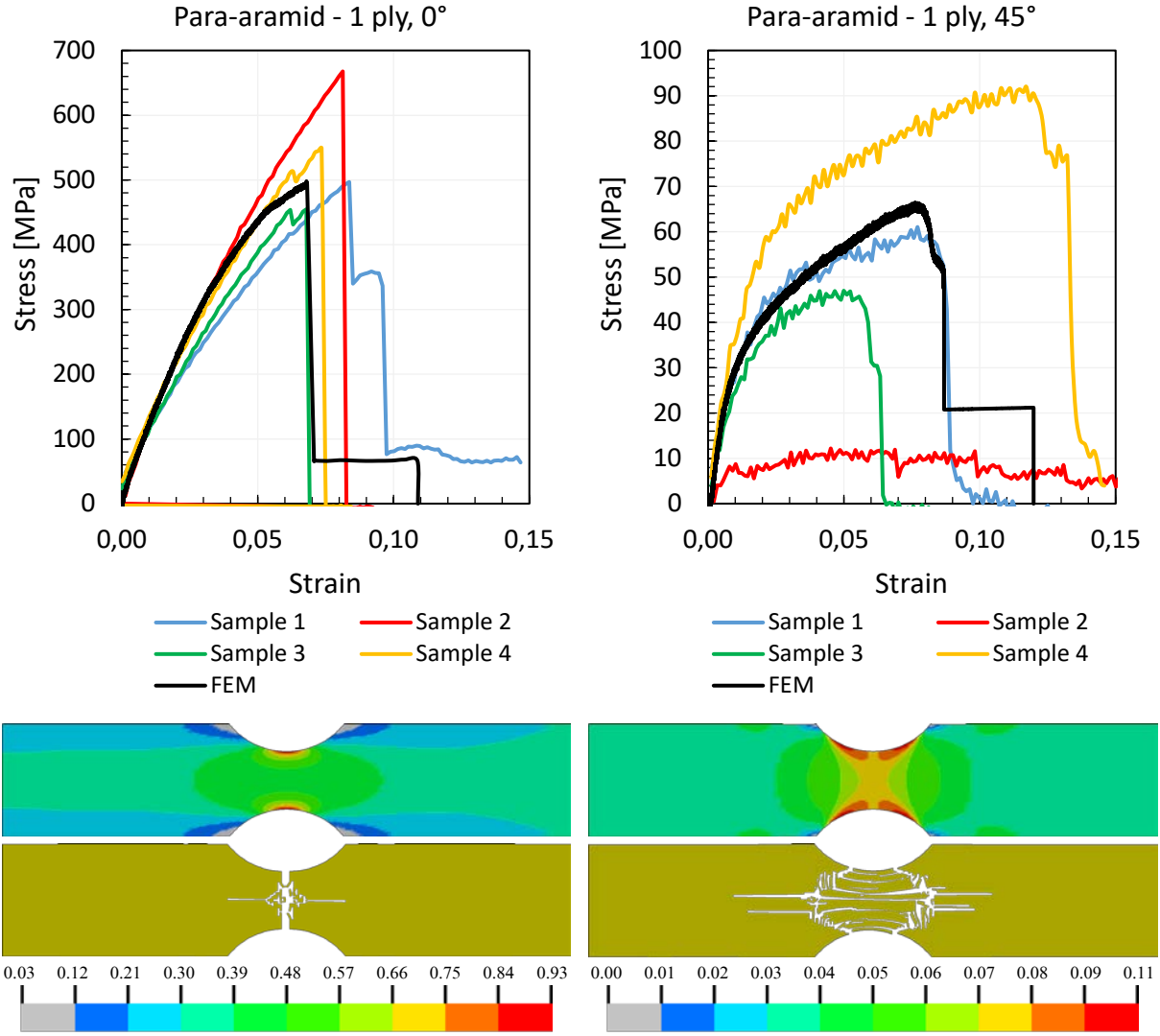

**Figure 6.** Experimental and FEM stress-strain curves for 1 ply of PA laminate at 0° (left panels) and 45° (right panels) of orientation of the warp with respect to the direction of application of the load (horizontal direction, see Figure S2 in the Supplementary Information). Fibre volume fractions are 0.533 and 0.400 respectively. The bottom panel shows, for the two orientations, the contour plot of von-Mises stress (in GPa) at the failure onset and the images of the failed samples (eroded elements) as obtained from FEM simulations.

**Table 4.** Average laminate tensile strength (and related standard deviation) from experimental data and comparison with FEM simulation results (values extracted from Figure 6 and Figures S5-S12 in the Supplementary Information) and prediction from rule of mixture (Equation 2).

| | 0° | | | | | | | | |
|---|---|---|---|---|---|---|---|---|---|
| | 1 layer | | | 5 layers | | | 10 layers | | |
| | Exp. | FEM | Eq. 2 | Exp. | FEM | Eq. 2 | Exp. | FEM | Eq. 2 |
| **Carbon** | 367.61±99.10 | 498.00 | 979.41 | 533.16±42.44 | 485.56 | 1065.29 | 515.40±60.78 | 473.10 | 1004.61 |
| **E-glass** | 208.19±6.15 | 249.00 | 474.93 | 329.75±48.93 | 348.61 | 670.81 | 308.87±11.55 | 344.18 | 696.10 |
| **PA** | 542.18±79.80 | 498.02 | 1377.74 | 460.25±34.09 | 560.26 | 1660.00 | 308.40±16.14 | 401.38 | 1191.25 |
| | 45° | | | | | | | | |
| | 1 layer | | | 5 layers | | | 10 layers | | |
| | Exp. | FEM | Eq. 2 | Exp. | FEM | Eq. 2 | Exp. | FEM | Eq. 2 |
| **Carbon** | 76.30±10.40 | 122.60 | 480.78 | 107.60±11.30 | 110.71 | 546.08 | 79.87±20.51 | 99.64 | 589.42 |
| **E-glass** | 50.01±27.69 | 77.50 | 229.30 | 50.68±11.55 | 66.43 | 284.90 | 54.80±13.59 | 71.96 | 354.59 |
| **PA** | 66.67±18.81 | 66.42 | 547.38 | 49.34±5.88 | 66.92 | 886.72 | 32.93±13.90 | 42.31 | 660.03 |

where the subscript $x$ represents the loading direction of the composite and $y$ its orthogonal, the subscripts 1 and 2 indicate the mutually orthogonal directions of the warp and weft of the woven textile (see Figure S2 in the Supplementary Information for the notation of quantities). Note that corresponding mechanical properties for both fibre (bi-directional textiles) and matrix (isotropic material) are equal in our case (i.e., $\sigma_1 = \sigma_2$) and that we have cautelatively assumed for the woven $\sigma_{12} = 0$, being negligible with respect to the corresponding counterpart in the principal direction. Results from Equation 2 (fracture strength $\sigma_{c,x} = \sigma_{c,y}$) are reported in Table 4. It is evident how the rule of mixtures significantly overestimates the properties of the composite for both orientations of the laminae with respect to the applied load. Alternatively, by back calculating the textile strength using Equation 2, we obtain significantly smaller values than those actually measured for the single bundle, showing, as expected, size scale effects on material properties (Figure S13 in the Supplementary Information). Thus, experiments and simulations are necessary complementary tools to characterize the material at the laminate level and predict accurate values of the fracture strength.

Analysing the stress-strain curves reported in Figure 6 (and Figures S5-S12 in the Supplementary Information) we observe that under tensile load, in general, a first sublinear phase is present, during which there is simultaneous matrix fracture, fibre debonding and fracture in the loading direction, up to the maximum load [4]. This progressive failure and softening is also predicted by FEM where all these mechanism cannot be accounted for, but this behaviour derives from the delayed reaching of the post peak phase and overcoming of the failure criterion at each integration points through the thickness of the thick shell elements. Subsequently, there is an unloading phase with residual effects due to frictional sliding of the reinforcing fibres in the matrix and residual matrix strength up to final fracture.

It can be noticed that when loading is applied at a 45° angle with respect to the fibre direction ($\theta$), there is a greater variability in the results for stress/strain curves: this is due to the greater sensitivity with respect to geometrical (i.e., fabrication) imperfections and the consequent variability in determining the onset and propagation of damage , i.e. the post-peak stress-strain behavior. In this case, experimental curves have a common initial slope (i.e. Young's modulus), but vary considerably in the damage evolution part of the curve. Despite this, the FEM simulations correctly reproduce the average experimental behaviour, in terms of average strength, ultimate strain, and specific toughness values.

*3.3 <u>FEM impact simulations</u>*

Figure 7 reports the results of FEM impact simulations, in terms of evolution of the projectile translational velocity vs. time for the four armours analysed under ballistic tests. The projectile residual velocities after impact ($V_{res}$) predicted by FEM simulations are 103 m/s, 115 m/s, 0 m/s (stopped projectile), and 3 m/s for tests on carbon, E-glass, PA ($t$ = 5 mm, and $t$ = 7 mm, respectively). The corresponding experimental values [47] are 110 m/s, 110 m/s, 27 m/s, and 0 m/s, respectively. Note that in the case of PA armor, being the scenario near to the critical limit (i.e., $V_{50}$), the difference in the occurrence of perforation between experiments and simulations falls within the statistical variation and model uncertainty. Figure 8 provides a visual comparison between the damage distribution in simulated plates and experiments, showing good agreement in the deformation behaviour. Thus, the developed numerical model, based on the mechanical properties of each single component, is able to predict with a good level of reliability the energy absorption capability of the targets, related damage and failure mechanisms. It is also verified that in the velocity regime analysed in this work the consideration of strain-rate effect is not significant even in absolute sense (see Figure S14 in the Supplementary Information).

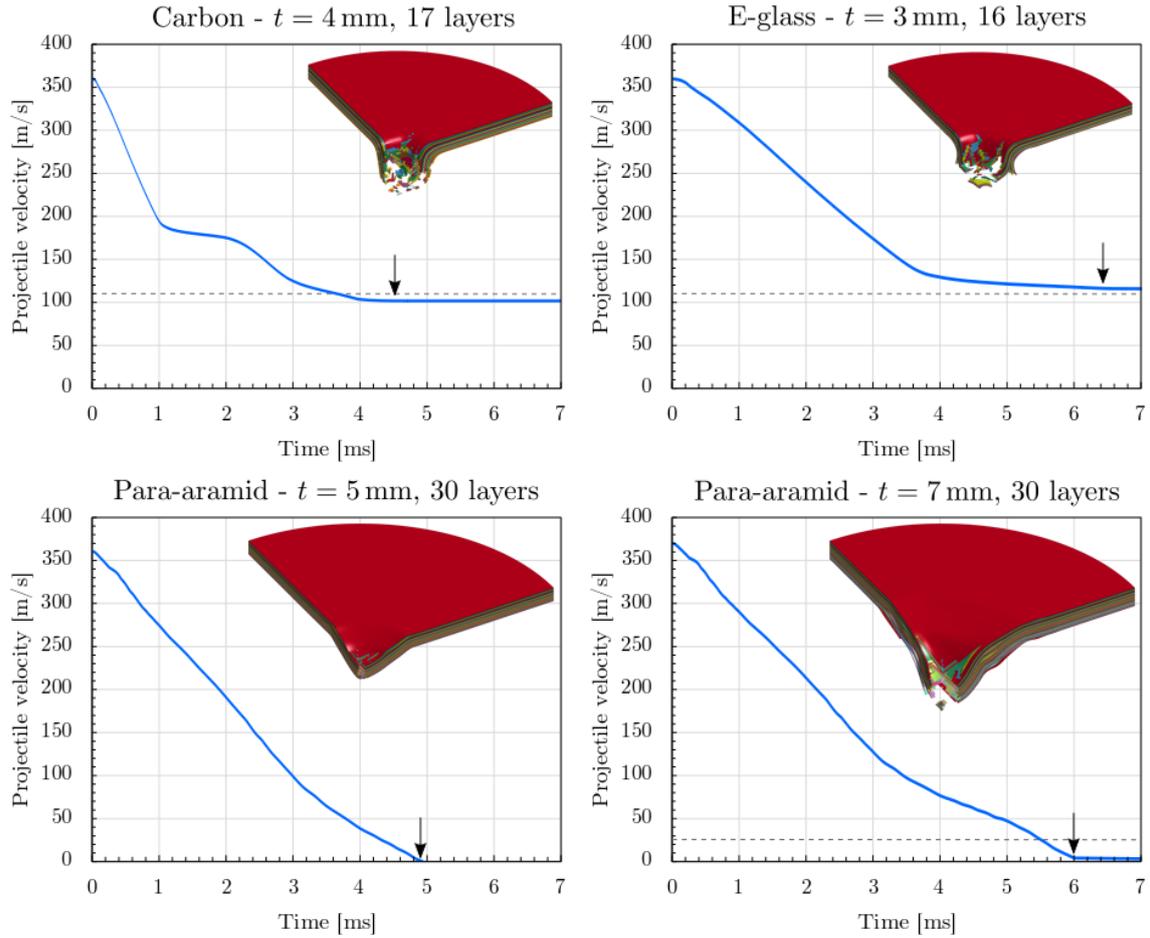

**Figure 7.** Evolution of the projectile velocity over time after impact with the four tested targets by FEM simulation. The dashed lines represent the reference value of the residual velocity determined from ballistic experiments [47] (for the PA plate with $t = 5$ mm both simulation and experiment provide $V_{res} = 0$). The insets depict the snapshots of FEM simulations taken at the time at which the projectile velocity stabilizes after the strike, either $V_{res} > 0$ or the projectile is stopped, and highlighted by the arrows on the curves.

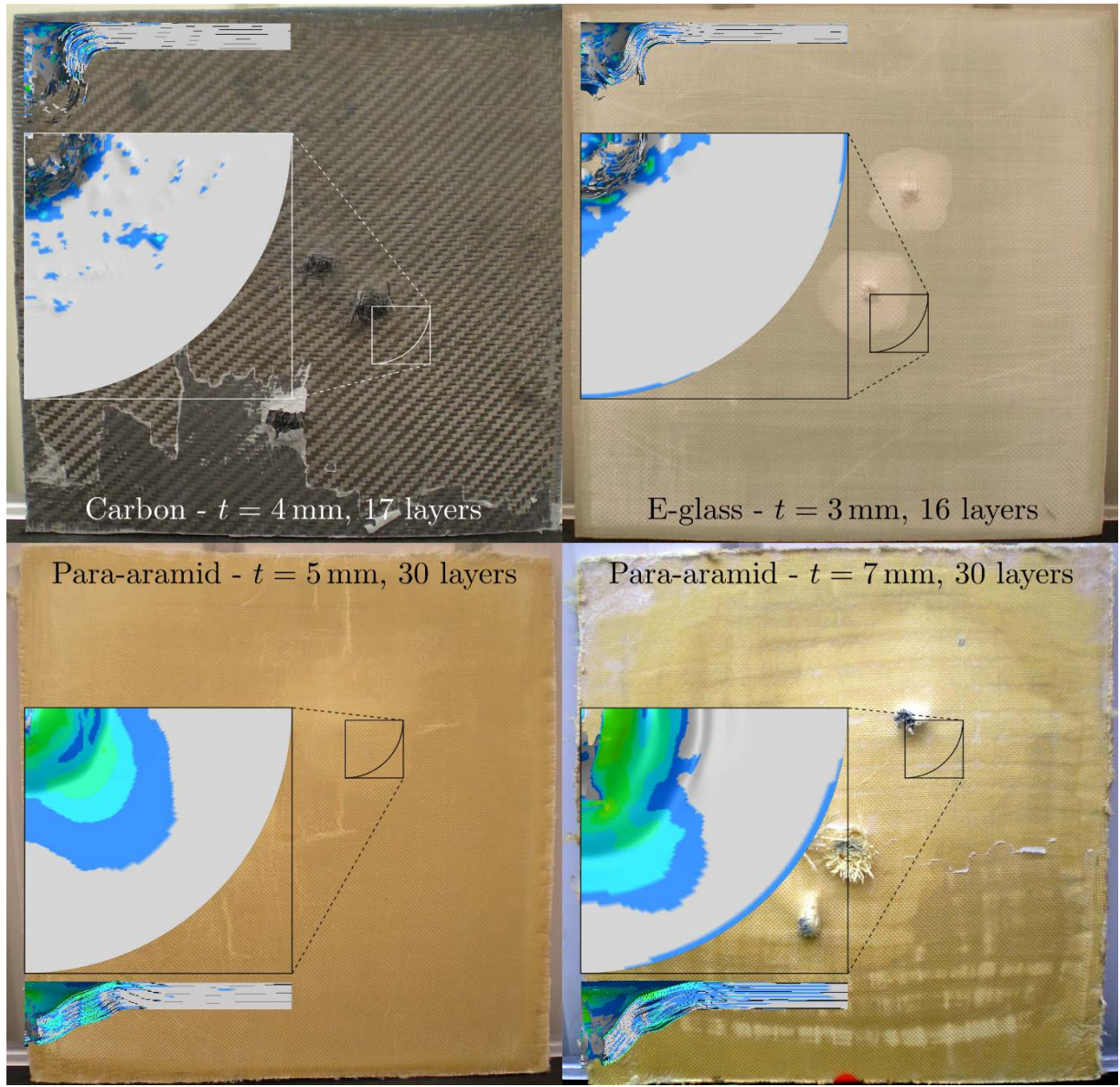

**Figure 8.** Visual comparison after impact at $V_{\text{res}} = 360$ m/s between experimental (rear face) and simulated targets (rear face and cross section). The magnified regions have a size of 40x40 mm² (overall size of the experimental target is ~370x370 mm²) and refers to the first impact performed on the armour. The stresses in the FEM images (von-Mises) are plotted to highlight qualitatively the radius of the zone affected by the impact and compare it with the deformation observed in experiments. Experimental pictures courtesy of Vemar Helmets s.r.l.

As expected, high strength fibres with limited toughness due to low ultimate strain (carbon) or low strength (glass) display a more localized damage and, consequently, their absolute and specific impact toughness is smaller with respect to PA plates. On the contrary, PA plates are able to undergo larger and less localized deflection and deformation, also promoting delamination over a wider area, giving a more synergistic contribution of energy dissipation between the layers [12]. This translates overall into higher impact energy absorption capability. However, a primary requirement in ballistic applications, especially for body armours, is to minimize the target perforation depth and deformation: in this sense, a good balance between strength and ultimate strain to failure is necessary to maximize the toughness –or to avoid its impairment– within given deformability constraints. Results are in agreement with observation at lower impact velocities [33].

From the comparison of the two PA plates, it is possible to notice the effect of the composite volume fraction, derived from different curing pressures and temperatures, which allows the thinner 5 mm plate to stop the projectile in a shorter time (and thickness) providing a higher specific energy absorption capability (energy per layer or per areal density) with respect to the 7 mm thickness counterpart. This aspect is not predicted by the classical dimensional analysis [11]. To rationalize this latter result and evaluate and compare the energy absorption capability of the three materials when used as reinforcement in armours, we propose a multiscale generalization to heterogeneous materials of the Cuniff's parameter, originally developed for plain textiles, by taking into account the composite nature of the target, as follows:

$$U_\text{m} \sim \frac{[f\sigma_\text{bundle} + (1-f)\sigma_\text{m}]^2}{2[f\rho_\text{bundle} + (1-f)\rho_\text{m}][fE_\text{bundle} + (1-f)E_\text{m}]} \sqrt{\frac{E_\text{bundle}}{\rho_\text{bundle}}} \qquad (3)$$

where the properties of the bundle, which can be in turn inferred by the properties of the single fibres through Equation 1, are explicitly considered. Note that the composite material strain here is calculated as $\varepsilon_\text{c} = \frac{\sigma_\text{c}}{E_\text{c}} = \frac{[f\sigma_\text{bundle}+(1-f)\sigma_\text{m}]}{[fE_\text{bundle}+(1-f)E_\text{m}]}$, while the term under the square root related to the dissipation by elastic waves accounts only for the reinforcement phase since the elastic wave will be guided in the plane primarily within the stiffer phase of the composite, i.e. the textile, regardless of the volume fraction $f$.

Results scaled according to Equation 3 are reported in Figure 9, allowing to compare on the same graph the performance of different reinforcing materials also structured in the composite in different ways (volume fraction and number of layers). It is then possible to make a more realistic comparison among materials, for examples taking into account the issues that some textile or mould geometries may create in obtaining desired volume fraction, due to specific difficulties in the production process [48]. The good correlation between the lower scale material parameters (input) derived experimentally and the performance of the armour extracted from ballistic impact simulations (output) shows the good capability of the modified criterion, as well as the inferring of properties from single constituents, to predict the impact performance at the macro scale, starting from the properties of the constituents at the micro-scale. Thus, the proposed multiscale characterizationprocess, summarized in Figure 10, can provide a preliminary and effective assessment of the suitability of different reinforcing materials and the selection of optimal ones for impact energy absorption and shielding.

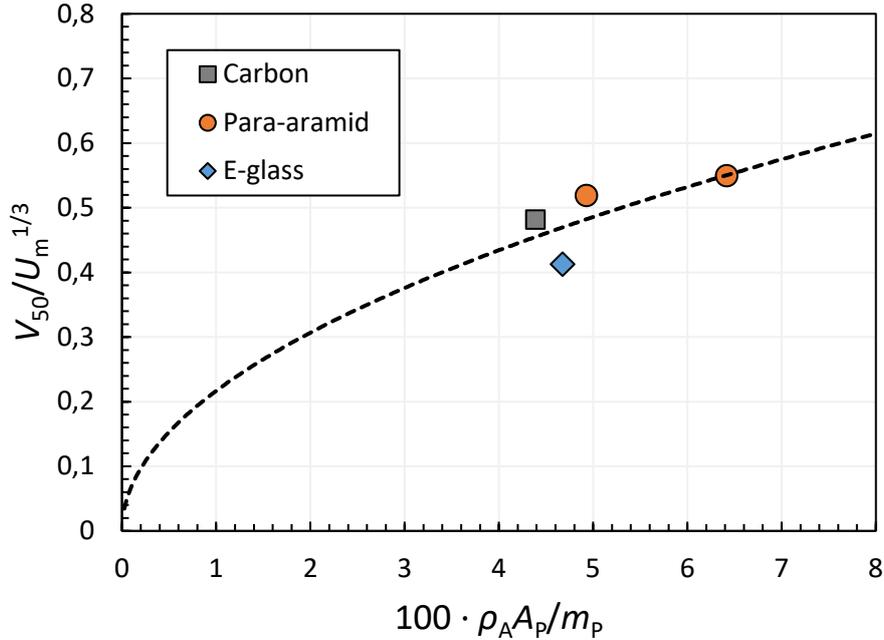

**Figure 9.** Comparison on the Cuniff's map of the three tested materials and four armours structures by Equation 3. The ballistic limit velocity $V_{50}$ is extracted from FEM simulations and corresponds, in this case, to the condition $V_{res} = 0$. $\rho_A = \rho t$ is the areal density of the target while $A_P = \pi \Phi^2/4$ is the projected area of the projectile with mass $m_P$ (projectile is the same in all cases).

## 4. Conclusions

In this paper we proposed a multiscale coupled theoretical/experimental/numerical framework to provide consistent and reliable correlation between of tensile (quasi-static) and impact properties of composite laminates. Starting from the characterization of the single fibres using a nanotensile testing machine and of fibre bundles at mesoscale, we used the measured tensile properties as input of non-linear FEM model to predict the tensile fracture properties of the laminates at macroscale, verifying them with experimental results. Then, a multilayer integration model for single plies was assembled and employed to construct, via the introduction of contact algorithms and proper boundary conditions, a numerical model of multilayer armours subjected to high-velocity impact, whose predictions were verified with ballistic tests. The predictions of impact energy absorption obtained by using microscale properties are in good agreement with experimental results, additionally showing a direct correlation between the fibre properties and their structural arrangement (in terms of volume fraction) with the limit ballistic velocity, by employing a proposed multiscale generalization of the Cuniff's parameter. We have thus demonstrated that a characterization of the mechanical properties via simple tensile tests can help to preliminarily assess and compare the suitability of different materials for employment as reinforcement in composite armours for ballistic application. The multiscale characterization presented in this work can allow to extend traditional design concepts of composites for ballistic applications to novel nanofibres and nanocomposites [49, 50], with the potential capability to also integrate the role of hierarchical structures and geometries at multiple levels.

**Acknowledgements**

NMP is supported by the European Commission under the Graphene Flagship Core 2 grant No. 785219 (WP14 "Composites") and FET Proactive "Neurofibres" grant No. 732344 as well as by the Italian Ministry of Education, University and Research (MIUR) under the "Departments of Excellence" grant L. 232/2016, the ARS01-01384-PROSCAN Grant and the PRIN-20177TTP3S Grant. FB is supported by H2020 FET Proactive Neurofibres Grant No. 732344, by project Metapp (n. CSTO160004) co-funded by Fondazione San Paolo, and by the Italian Ministry of Education, University and Research (MIUR) under the "Departments of Excellence" grant L. 232/2016. SHR is supported by the Basic Science Research Program (2019R1A2C4070690) and the Creative Materials Discovery Program (2016M3D1A1900038) through the National Research Foundation of Korea (NRF). SS acknowledges financial support from Brain Korea 21 Plus Postdoc Scholarship (NRF).

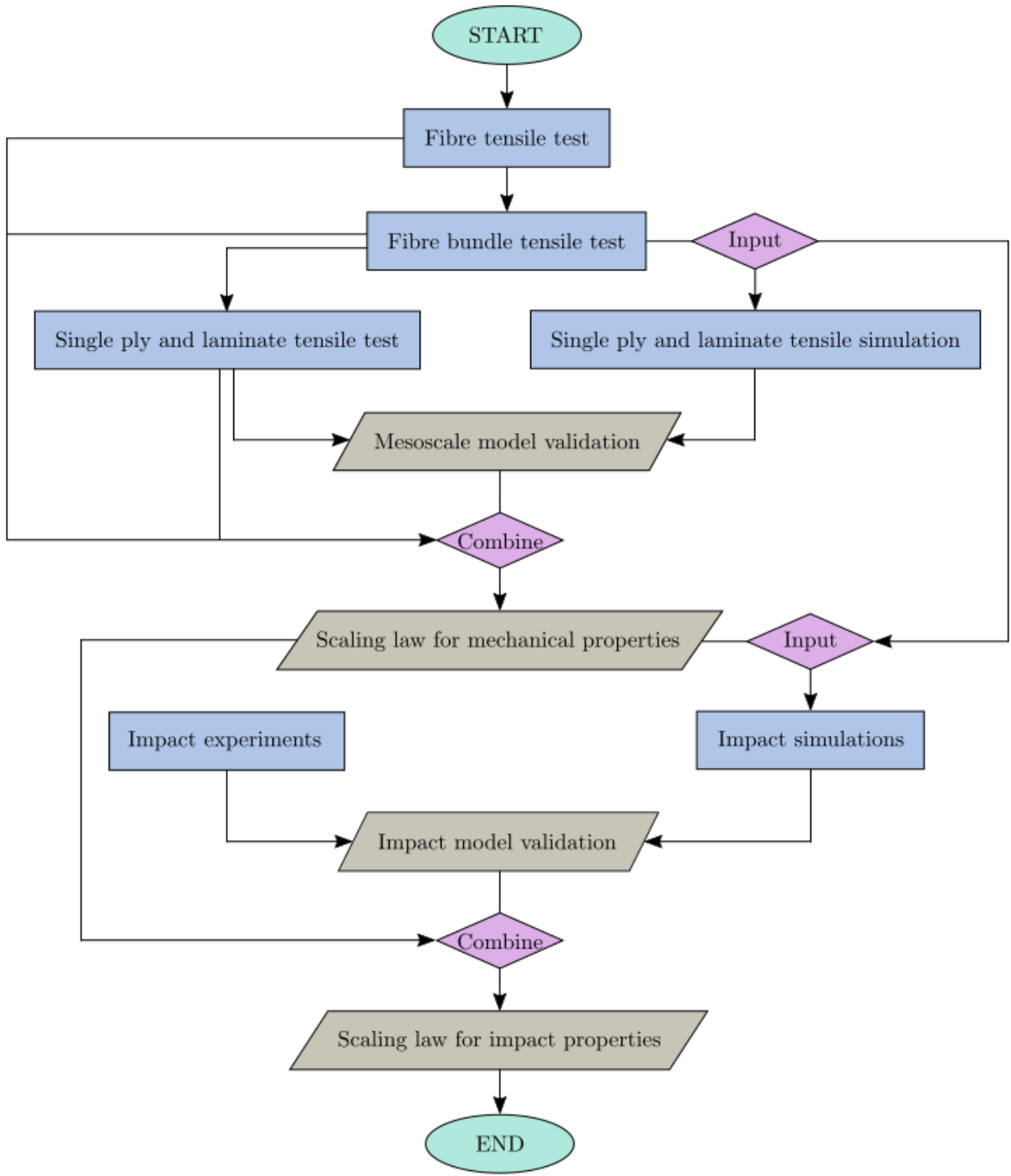

**Figure 10.** Flow chart of the proposed experimental-numerical approach for the scaling assessment of material properties for impact studies.